\documentclass[10pt,aps,prd,preprintnumbers,showpacs,nofootinbib,superscriptaddress,notitlepage,twocolumn,longbibliography]{revtex4-2}

\usepackage{graphicx}
\usepackage{xcolor}
\usepackage{amsmath}
\usepackage{amssymb}
\usepackage{enumerate}
\usepackage{listings}
\usepackage{multirow}
\usepackage{csvsimple}

\usepackage{array}
\newcolumntype{P}[1]{>{\centering\arraybackslash}p{#1}}
\newcolumntype{M}[1]{>{\centering\arraybackslash}m{#1}}

\usepackage[colorlinks=true,citecolor=blue,urlcolor=blue]{hyperref}
\usepackage{notes2bib}

\newcommand{\PRE}[1]{{#1}} 

\def\be{\begin{equation}}
\def\ee{\end{equation}}
\def\bea{\begin{eqnarray}}
\def\eea{\end{eqnarray}}

\newcommand{\new}[1]{#1}
\newcommand{\renew}[1]{#1}

\graphicspath{{./Results/2x2_of_all_profiles/}}
\begin{document}

\title{\texorpdfstring{$J$}{J}-factors for Velocity-dependent Dark Matter
Annihilation}

\author{Bradly Boucher}
\affiliation{\mbox{Department of Physics \& Astronomy,
University of Hawai`i, Honolulu, Hawai`i 96822, USA}}

\author{Jason Kumar}
\affiliation{\mbox{Department of Physics \& Astronomy,
University of Hawai`i, Honolulu, Hawai`i 96822, USA}}

\author{Van B. Le}
\affiliation{\mbox{Department of Physics \& Astronomy,
University of Hawai`i, Honolulu, Hawai`i 96822, USA}}

\author{Jack Runburg}
\affiliation{\mbox{Department of Physics \& Astronomy,
University of Hawai`i, Honolulu, Hawai`i 96822, USA}}

\begin{abstract}
\PRE{\vspace*{.1in}}
If dark matter annihilates with a velocity-dependent cross section within a subhalo, then the magnitude and angular distribution of the resulting photon signal will change.
These effects are encoded in the $J$-factor.
In this work we compute the $J$-factor for a variety of choices for  the cross section velocity-dependence, and for a variety of choices for the dark matter profile, including generalized Navarro-Frenk-White (NFW), Einasto, Burkert and Moore.
\new{
We include the results of these computations as data products alongside the article.
We find that  the angular distribution of a future signal 
would depend on the velocity-dependence of the annihilation cross section more strongly for cuspy profiles than for 
cored profiles.  
}
Interestingly, we find that for a density profile with an inner slope power law steeper than 4/3, Sommerfeld-enhanced annihilation in the Coulomb limit leads to a divergence at the center, requiring a more detailed treatment of departure from the Coulomb limit.
\end{abstract}

\maketitle
\section{Introduction}

A promising strategy for the indirect detection of dark matter is the search for photons arising from dark matter annihilation in galactic subhalos, including those which host dwarf spheroidal galaxies 
(dSphs)~\cite{Fermi-LAT:2015att,Geringer-Sameth:2014qqa,MAGIC:2016xys,HAWC:2017mfa}.
This strategy is promising because the photons will point back to the subhalo, which is a region with a large dark matter density, but relatively small baryonic density~\cite{Mateo_1998,McConnachie_2012}.
There is thus relatively little astrophysical fore/background to a potential dark matter signal.
The dependence of this photon signal on the properties of an individual subhalo is encoded in the $J$-factor, which in turn depends on the dark matter velocity distribution in the subhalo, and on the velocity-dependence of the dark matter annihilation cross section.

Different models for the velocity-dependence of the dark matter annihilation cross section can lead to $J$-factors with different normalizations and angular dependences~~\cite{Robertson:2009bh,Belotsky:2014doa,Ferrer:2013cla,Boddy:2017vpe,Zhao:2017dln,Petac:2018gue,Boddy:2018ike,Lacroix:2018qqh,Boddy:2019wfg,Boddy:2020,Ando:2021jvn}.
In this way, the microphysics of the dark matter annihilation cross section is connected to both the amplitude and morphology of the resulting photon signal.
For this reason, it is important to determine $J$-factors which arise under all theoretically-motivated assumptions for the velocity-dependence of the cross section.
The most well-studied case is $s$-wave annihilation, in which $\sigma v$ is velocity-independent.
In recent work \citep[e.g.,][]{bergstrom:2018asd}, $J$-factors have been calculated for other well-motivated examples, such as $p$-wave, $d$-wave, and Sommerfeld-enhanced annihilation.
But most of these calculations have been performed under the assumption that the dark matter density profile $\rho(r)$ is of the Navarro-Frenk-White (NFW) form~\cite{Navarro:1995iw}.
Our goal in this work is to generalize this calculation to other density profiles which are commonly used, and motivated by $N$-body simulation results.

We will consider generalized NFW, Einasto~\cite{Einasto:1965czb}, Burkert~\cite{Burkert:1995yz}, and Moore~\cite{Moore:1997sg} profiles.
Like the standard NFW profile, these density distributions are characterized by only two dimensional parameters, $\rho_s$ and $r_s$.
The dependence of the $J$-factor on these parameters is largely determined by dimensional analysis~\cite{Boddy:2019wfg}.
Given our results, one can easily determine the amplitude and angular distribution of the photon signal for any subhalo and choice of density profile, in terms of the halo parameters and the velocity-dependent cross section.

Our strategy will be to use the Eddington inversion method~\cite{10.1093/mnras/76.7.572} to determine the dark matter velocity distribution $f(r,v)$ from $\rho(r)$.
This velocity-distribution will, in turn, determine the $J$-factor.
For each functional form, we will be able to determine a scale-free $J$-factor which depends on velocity-dependence of the annihilation cross section, but is independent of the halo parameters.
The dependence of the $J$-factor on $\rho_s$ and $r_s$ is entirely determined by dimensional analysis.
This will leave us with a set of dimensionless numerical integrals to perform, for any choice of the velocity-dependence and of the density distribution functional form, which in turn determine the $J$-factor for any values of the subhalo parameters.  

\renew{We will also find that, for some classes of 
profiles, one can find analytic approximations for the velocity
and angular distributions.  These analytic computations will yield 
insights which generalize to larger classes of profiles than those 
we consider.  For example, we will find that, in the case of 
Sommerfeld-enhanced annihilation, the annihilation rate has a 
physical divergence if the inner slope of the profile 
is steeper than $4/3$ (independent of the shape at large distance), 
requiring one to account for deviations 
from the Coulomb limit.}

The plan of this paper is as follows.  In Section~\ref{sec:formalism}, we review the 
general formalism for determining the $J$-factor.  In Section~\ref{sec:models}, we describe 
the models of dark matter particle physics and astrophysics which we will consider.  
We present our results in Section~\ref{sec:results}, and conclude in Section~\ref{sec:conclusion}.

\section{General Formalism}
\label{sec:formalism}

We will follow the formalism of~\cite{Boddy:2019wfg}, which we review here.
We consider the scenario in which the dark matter is a real particle whose annihilation cross section can be approximated as $\sigma v = (\sigma v)_0 \times S(v/c)$, where $(\sigma v)_0$ is a constant, independent of the relative velocity $v$.

The $J$-factor describes the astrophysical contribution to the dark matter annihilation flux
\bea
J_S (\theta) &=& \int d\ell \int d^3 v_1 \int d^3 v_2~ 
f({\bf r}(\ell, \theta), {\bf v}_1)~ 
f({\bf r}(\ell, \theta), {\bf v}_2)
\nonumber\\
&\,& \times
S(|{\bf v}_1 - {\bf v}_2|/c) ,
\label{eq:JFactor}
\eea
where $f$ is the dark matter velocity distribution,     $\ell$ is the distance along the line of sight, and $\theta$ is the angle between the the line-of-sight direction and the direction from the observer to the center of the subhalo.  



\subsection{Scale-free \texorpdfstring{$J$}{J}}

We will assume that the dark matter density profile $\rho(r)$ depends only on two dimensionful parameters, $\rho_s$ and $r_s$.
In that case, we may rewrite the density profile in the scale-free form $\tilde \rho (\tilde r)$, where 
\bea
\tilde r &\equiv&   r / r_s ,
\nonumber\\
\tilde \rho (\tilde r) &\equiv& \rho (r) / \rho_s .
\eea
$\tilde \rho (\tilde r)$ has no dependence on the parameters $\rho_s$ and $r_s$.
Aside from $\rho_s$ and $r_s$, the only relevant dimensionful constant is $G_N$.
We also define a scale-free velocity using the only combination of these parameters with units of velocity,
\bea
\tilde v &\equiv& v / \sqrt{4\pi G_N \rho_s r_s^2} ,
\eea
in terms of which we may define the scale-free velocity distribution 
\bea
\tilde f (\tilde r, \tilde v) &\equiv& 
\left(4\pi G_N \rho_s r_s^2 \right)^{3/2} \rho_s^{-1} 
f (r,v) ,
\eea
where $\tilde \rho (\tilde r) = \int d^3 \tilde v ~ \tilde f (\tilde r, \tilde v)$ and where $\tilde f (\tilde r, \tilde v)$ is independent of the dimensionful parameters.

We will assume that the velocity-dependence of the dark matter annihilation cross section has a power-law form, given by $S(v/c) = (v/c)^n$.
We may then express the $J$-factor in scale-free form.
\bea
J_{S(n)} (\tilde \theta) &=& 2 \rho_s^2 r_s 
\left(\frac{4\pi G_N \rho_s r_s^2}{c^2} \right)^{n/2} 
\tilde J (\tilde \theta) ,
\nonumber\\
J_{S(n)}^{tot}  &=& \frac{4\pi \rho_s^2 r_s^3}{D^2} 
\left(\frac{4\pi G_N \rho_s r_s^2}{c^2} \right)^{n/2} 
\tilde J^{\rm tot} ,
\label{eq:JFactorToScaleFree}
\eea
where the scale-free quantities $\tilde J_{S(n)} (\tilde \theta)$ and $\tilde J_{S(n)}^{tot}$ are given by~\cite{Boddy:2019wfg}
\bea
\tilde J_{S(n)}^{tot} &\approx& 
\int_0^\infty d\tilde \theta ~ \tilde \theta ~ 
\tilde J_{S(n)} (\tilde \theta) ,
\nonumber\\
\tilde J_{S(n)} (\tilde \theta) &\approx& 
\int_{\tilde \theta}^\infty d\tilde r ~
\left[1 - \left(\frac{\tilde \theta}{\tilde r}\right)^2 
\right]^{-1/2} P_n^2 (\tilde r),
\label{eq:Js_Jstot}
\eea
and where 
\bea
P_n^2 &=& \int d^3 \tilde v_1 d^3 \tilde v_2 ~
|\tilde {\bf v}_1 - \tilde {\bf v}_2|^n
\tilde f (\tilde r, \tilde v_1)
\tilde f (\tilde r, \tilde v_2) .
\label{eq:P2n}
\eea
In the case of $s$-wave annihilation, $P_{n=0}^2 = \tilde \rho^2$.  $P_n^2$ is thus 
the generalization of $\tilde \rho^2$ relevant to computation of the $J$-factor for velocity-dependent dark matter annihilation.

Note, that if $n$ is a positive, integer, then 
the expression for $P_n^2$ can be expressed in terms 
of one-dimensional integrals.  In particular, we 
find
\bea
P_{n=2}^2 (\tilde r) &=& 
[\tilde \rho (\tilde r)]^2 \left[
2\langle \tilde v^2 \rangle  (\tilde r)\right] ,
\nonumber\\
P_{n=4}^2 (\tilde r) &=& 
[\tilde \rho (\tilde r)]^2 \left[
2 \langle \tilde v^4 \rangle (\tilde r)
+ \frac{10}{3} \left(\langle \tilde v^2 \rangle  (\tilde r) \right)^2
\right]
\eea
where $\langle \tilde v^m \rangle (\tilde r) 
= 4\pi [\int_0^\infty d\tilde v ~ \tilde v^{m+2} 
\tilde f (\tilde r, \tilde v)] / \tilde \rho 
(\tilde r)$.
For the case of $n=-1$, one must perform the two-dimensional integral.

\subsection{Eddington Inversion}

If the subhalo is in equilibrium, then the velocity-distribution can be written as a function of the integrals of motion.
Since we have assumed that the velocity distribution is spherically symmetric and isotropic, it can be written as a function only of the energy per particle, $E = v^2/2 + \Phi(r)$, where $\Phi(r)$ is the gravitational potential\footnote{Following convention, we use the symbol $\Phi$ for both the photon flux and the gravitational potential.  We trust the meaning of $\Phi$ will be clear from context.} (that is, $f(r,v) = f(E(r,v))$).
The velocity distribution can then be expressed in terms of the density using the Eddington inversion formula~~\cite{10.1093/mnras/76.7.572}, yielding
\bea
f(E) &=& \frac{1}{\sqrt{8} \pi^2 } \int_E^{\Phi(\infty)} 
\frac{d^2 \rho}{d \Phi^2} \frac{d\Phi}{\sqrt{\Phi - E}} ,
\eea
where 
\bea
\Phi(r) &=& \Phi(r_0) + 4\pi G_N \rho_s r_s^2 
\int_{\tilde r_0}^{\tilde r} \frac{dx}{x^2} 
\int_0^x dy ~y^2 \tilde \rho(y) .
\eea
Note, we have assumed that the baryonic contribution to the gravitational potential is negligible.

In terms of the scale-free gravitational potential and energy $\tilde \Phi (\tilde r) \equiv \Phi (r) / 4\pi G_N \rho_s r_s^2$, $\tilde E \equiv E  / 4\pi G_N \rho_s r_s^2$, we then find
\bea
\tilde f(\tilde r, \tilde v) = 
\tilde f(\tilde E(\tilde r, \tilde v)) 
&=& \frac{1}{\sqrt{8} \pi^2 } \int_{\tilde E}^{\tilde \Phi(\infty)} 
\frac{d^2 \tilde \rho}{d \tilde \Phi^2} \frac{d\tilde \Phi}{\sqrt{\tilde \Phi - \tilde E}} .
\nonumber\\
\label{eq:ScaleFreeEddignton}
\eea

The scale-free quantities $\tilde J$ and $\tilde J^{tot}$ depend on the functional form of the dark matter density distribution ($\tilde \rho$), and on the velocity dependence of the annihilation cross section ($n$), but are independent of the parameters $\rho_s$ and $r_s$.
For any functional form of $\tilde \rho$, and any choice of $n$, one can compute 
$\tilde J (\tilde \theta)$ and $\tilde J^{tot}$ by performing the integration described above.
For any individual subhalo with parameters $\rho_s$ and $r_s$, a distance $D$ away from Earth, the $J$-factor is then determined by Eq.~\ref{eq:JFactorToScaleFree}.
This calculation has been performed for the case of an NFW profile, in which case $\tilde \rho (\tilde r) = \tilde r^{-1} (1+ \tilde r)^{-2}$ \citep{Boddy:2019wfg}.
We will extend this result to a variety of other profiles.

\section{Dark Matter Astrophysics and Microphysics }
\label{sec:models}

We will consider four theoretically well-motivated scenarios for the power-law velocity dependence of the dark matter annihilation cross section ($S(v/c) = (v/c)^n$).
\begin{itemize}
\item{$n=0$ ({\it $s$-wave}): 
\new{In this case, the dark matter initial 
state has orbital angular momentum $L=0$, and }$\sigma v$ is independent of $v$ 
\new{in the non-relativisitic limit}.  This is the standard case which is 
usually considered.
}
\item{$n=2$ ({\it $p$-wave}): 
\new{In this case, the dark matter initial 
state has orbital angular momentum $L=1$.}
This case can arise if dark matter is a Majorana fermion which annihilates to a Standard Model (SM) fermion/anti-fermion 
pair through an interaction respecting minimal flavor violation (MFV) (see, for 
example,~\cite{Kumar:2013iva}). }
\item{$n=4$ ({\it $d$-wave}): 
\new{In this case, the dark matter initial 
state has orbital angular momentum $L=2$.}
This case can arise if dark matter is a real scalar annihilating to an SM fermion/anti-fermion pair through an interaction respecting MFV (see, for example,~\cite{Kumar:2013iva,Giacchino:2013bta,Toma:2013bka}). }
\item{$n=-1$ ({\it Sommerfeld-enhancement in the Coulomb limit}):  This case can arise if there is a long-range attractive force between dark matter particles, mediated by a very light particle.
\new{If the dark matter initial state is $L=0$, a
$1/v$ enhancement arises because the dark matter initial 
state is an eigenstate of the Hamiltonian with a long-range 
attractive potential.  If the mediator has non-zero mass, then 
the $1/v$ enhancement will be cutoff for small enough velocity, but we focus on the case in which this cutoff is well below 
the velocity scale of the dark matter particles. For a detailed discussion, see~\cite{ArkaniHamed:2008qn,Feng:2010zp}, 
for example.}
}
\end{itemize}

\new{Despite significant effort, there is no 
consensus on the functional form of the dark matter 
profile which one should expected in subhalos.} 
We consider various dark matter profiles, which are motivated by $N$-body simulations \new{ and stellar 
observations}:
\begin{itemize}
\item{{\it Generalized NFW}
[$\tilde \rho(\tilde r) = \tilde r^{-\gamma} (1+\tilde r)^{-(3-\gamma)}$]:   
$\gamma=1$ corresponds to the standard NFW case
\new{\cite{Navarro:1995iw}, and was originally proposed as a good fit to the density found 
in $N$-body simulations.  
The generalization to $\gamma \neq 1$ was first studied in~\cite{Zhao:1996mr}, and has been argued to 
be a good fit $N$-body simulation results for larger values of $\gamma$~\cite{Klypin:2000hk}, 
although previous work had also indicated that smaller values of $\gamma$ may also be acceptable~\cite{Klypin:1997fb}.
We will consider a broad range of choices of $\gamma$ ranging from $0.6$ to $1.4$. 
(Note, for $\gamma \geq 1.5$, the $s$-wave annihilation rate would 
diverge.)}
}
\item \textit{Einasto profile}
[$\tilde \rho(\tilde r) = \exp (-(2/\alpha)(\tilde r^\alpha -1))$]:  
\new{This profile has been found to be at least 
as good 
fit as NFW to densities found in $N$-body simulations when $\alpha$ 
lies roughly in the range $0.12 < \alpha < 0.25$ 
(see, for example,~\cite{Gao:2007gh,Ludlow:2016qow}), 
and we will consider values of $\alpha$ in this range.}
\item{{\it Burkert profile} 
[$\tilde \rho(\tilde r) = (1+\tilde r)^{-1} (1+\tilde r^2)^{-1}$]:  
This is a commonly used example of a cored profile\new{, which was found to be a good 
fit to observations of stellar motions in dwarf galaxies~\cite{Burkert:1995yz}}.}
\item{{\it Moore profile}
[$\tilde \rho(\tilde r) = (\tilde r^{1.4} (1+\tilde r)^{1.4})^{-1}$]: This is an example of a very cuspy profile\new{, which was found to be a good fit to the $N$-body simulations considered in~\cite{Moore:1997sg,Klypin:2000hk}}.}
\end{itemize}

\section{Results}
\label{sec:results}

For any choice of $\tilde \rho (\tilde r)$ and of $n$, the $J$-factor is determined by three parameters ($\rho_s$, $r_s$ and $D$), and by a scale-free normalization ($\tilde J_{S(n)}^{\rm tot}$) and an angular distribution ($\tilde J_{S(n)} (\tilde \theta) / \tilde J_{S(n)}^{\rm tot}$), which must be determined by numerical integration.  
We can characterize the angular size of gamma-ray emission from a subhalo with the quantity $\langle \theta \rangle /\theta_0$, defined as
\bea
\frac{\langle \theta \rangle}{\theta_0} &\equiv& 
\frac{\int_0^\infty d\tilde \theta ~ \tilde \theta^2 
\tilde J_{S(n)}(\tilde \theta)}{\tilde 
J_{S(n)}^{\rm tot}} .
\label{eq:theta}
\eea

\begin{table*}
\centering
\begin{tabular}{|M{0.5cm}|M{0.9cm}|M{0.9cm}|M{0.9cm}|M{0.9cm}|M{0.8cm}|M{0.8cm}|M{0.8cm}|M{0.8cm}|M{0.8cm}|M{0.8cm}|M{0.8cm}|M{0.8cm}|M{0.8cm}|M{0.8cm}|M{0.8cm}|M{1.1cm}|M{0.9cm}|}
\hline\hline 
\multicolumn{18}{|c|}{$\tilde{J}_{S(n)}^{tot}$} \\ [0.5ex]
\hline\hline
& \multicolumn{10}{|c|}{NFW ($\gamma$)} & \multicolumn{5}{|c|}{Einasto ($\alpha$)} & Burkert & Moore\\ [0.5ex]
\hline
n  & 0.6   & 0.7   & 0.8   & 0.9   & 1.0  & 1.1  & 1.2  & 1.25 & 1.3  & 1.4  & 0.13 & 0.16 & 0.17 & 0.20 & 0.24  & \multicolumn{2}{|c|}{ }\\
\hline
\begin{filecontents*}{Results/tables/j_over_jtot.csv}
n,NFW-0.6,NFW-0.7,NFW-0.8,NFW-0.9,NFW-1.0,NFW-1.1,NFW-1.2,NFW-1.3,NFW-1.4,Einasto-0.13,Einasto-0.16,Einasto-0.17,Einasto-0.17,Einasto-0.20,Einasto-0.24,Burkert,Moore
-1,0.26,0.33,0.43,0.59,0.83,1.25,1.99,2.57,3.39,---,15,10.7,9.73,7.7,6.07,0.1,---
0,0.1,0.13,0.18,0.24,0.33,0.49,0.77,0.99,1.31,2.45,11.4,8.53,7.88,6.45,5.26,0.034,2.58
2,0.038,0.051,0.07,0.098,0.14,0.22,0.35,0.46,0.62,1.19,18.7,14.6,13.6,11.4,9.43,0.0069,1.37
4,0.024,0.034,0.05,0.075,0.12,0.19,0.33,0.45,0.62,1.28,62.5,48.9,45.6,37.9,31,0.0022,1.64
\end{filecontents*}
\csvreader[head to column names=false, late after line = \\\hline]{Results/tables/j_over_jtot.csv}{}{\csvlinetotablerow}
\end{tabular}
\caption{Numerical values for 
the scale-free normalization $\tilde{J}_{S(n)}^{tot}$ (defined in Eq.~\ref{eq:Js_Jstot}) for 
$n=-1, 0, 2,$ and $4$, where the profile is taken to be either generalized NFW (with $\gamma$ as listed), 
Einasto (with $\alpha$ as listed), Burkert, or Moore.
}
\label{table:Jtot} 
\end{table*}

\begin{table*}[t]
\centering 
\begin{tabular}{|M{0.5cm}|M{0.83cm}|M{0.83cm}|M{0.83cm}|M{0.83cm}|M{0.83cm}|M{0.83cm}|M{0.83cm}|M{0.83cm}|M{0.83cm}|M{0.83cm}|M{0.83cm}|M{0.83cm}|M{0.83cm}|M{0.83cm}|M{0.83cm}|M{1.1cm}|M{0.9cm}|}
\hline\hline 
\multicolumn{18}{|c|}{$\langle \theta \rangle / \theta_0$} \\ [0.5ex]
\hline\hline
& \multicolumn{10}{|c|}{NFW ($\gamma$)} & \multicolumn{5}{|c|}{Einasto ($\alpha$)} & Burkert & Moore\\ [0.5ex]
\hline
n  & 0.6  & 0.7  & 0.8  & 0.9  & 1.0  & 1.1  & 1.2  & 1.25 & 1.3  & 1.4  & 0.13 & 0.16 & 0.17 & 0.20 & 0.24 & \multicolumn{2}{|c|}{ }\\
\hline
\begin{filecontents*}{Results/tables/theta_over_theta0.csv}
n,NFW-0.6,NFW-0.7,NFW-0.8,NFW-0.9,NFW-1.0,NFW-1.1,NFW-1.2,NFW-1.3,NFW-1.4,Einasto-0.13,Einasto-0.16,Einasto-0.17,Einasto-0.17,Einasto-0.20,Einasto-0.24,Burkert,Moore
-1,0.65,0.57,0.48,0.4,0.32,0.24,0.18,0.15,0.12,---,0.18,0.23,0.24,0.29,0.33,0.53,---
0,0.71,0.63,0.55,0.47,0.39,0.32,0.25,0.21,0.18,0.12,0.24,0.29,0.3,0.34,0.38,0.51,0.15
2,0.73,0.66,0.59,0.52,0.45,0.38,0.31,0.28,0.24,0.18,0.31,0.35,0.37,0.39,0.42,0.47,0.22
4,0.73,0.66,0.59,0.52,0.46,0.39,0.33,0.3,0.26,0.2,0.34,0.38,0.39,0.41,0.43,0.44,0.26
\end{filecontents*}
\csvreader[head to column names=false, late after line = \\\hline]{Results/tables/theta_over_theta0.csv}{}{\csvlinetotablerow}
\end{tabular}
\caption{
Numerical values for the angular distribution
$\langle \theta \rangle / \theta_0$ (defined in Eq.~\ref{eq:theta}) for 
$n=-1, 0, 2,$ and $4$, where the profile is taken to be either generalized NFW (with $\gamma$ as listed), 
Einasto (with $\alpha$ as listed), Burkert, or Moore.
}
\label{table:angular} 
\end{table*}

\begin{figure*}[hptb]    
    \centering
    \includegraphics[width=\textwidth]{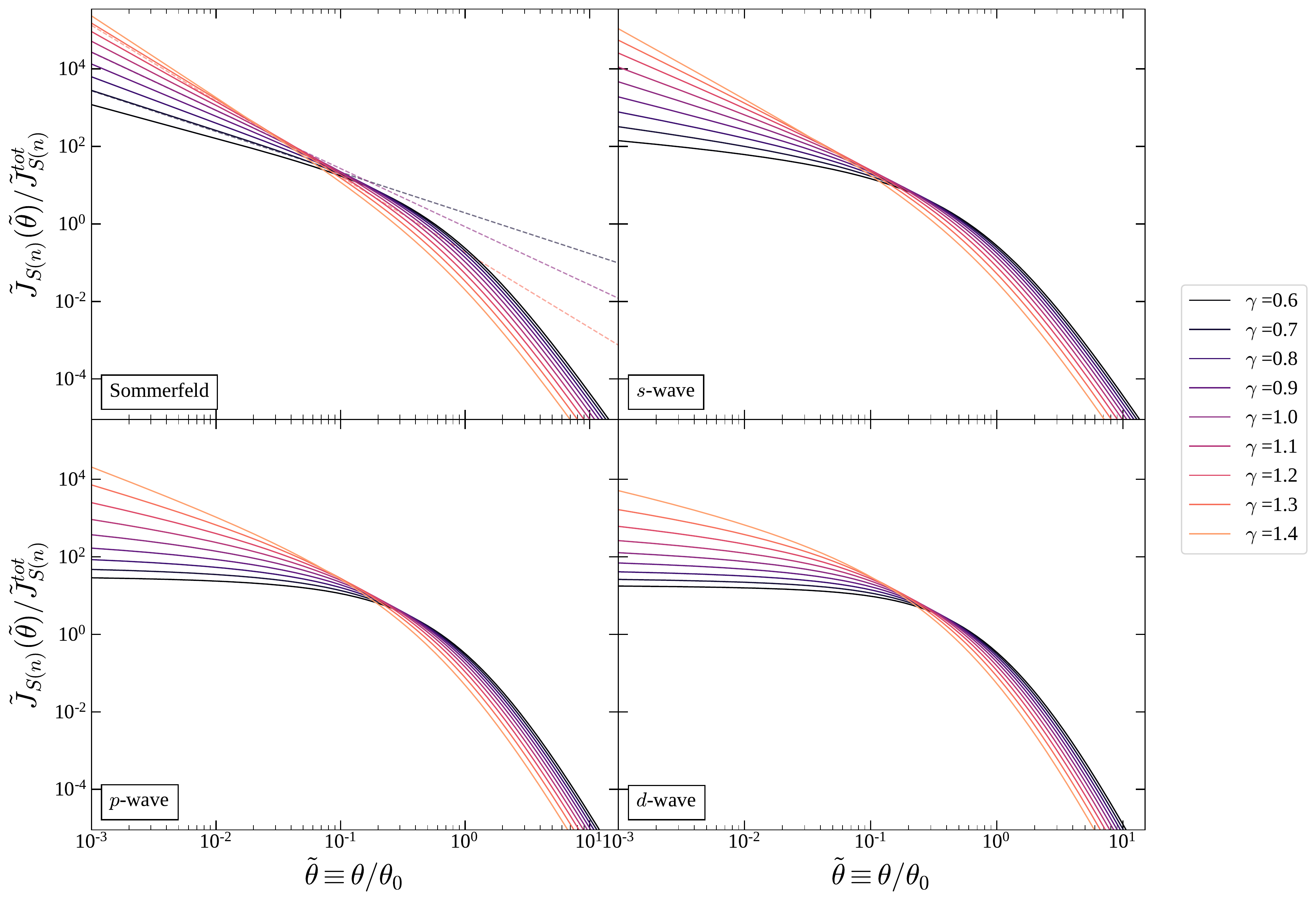}
    \caption{The scale-free photon angular distribution arising from Sommerfeld-enhanced \new{$n=-1$} (upper left), $s$-wave \new{$n=0$} (upper right), $p$-wave \new{$n=2$} (lower left), and $d$-wave \new{$n=4$} (lower right) dark matter annihilation in a generalized NFW subhalo \new{where the inner region goes like $\propto r^{-\gamma}$} (the profile parameter $\gamma$ varies from $0.6$ to $1.4$, as labelled). \new{The dashed lines show the analytic approximation from Eq.~\ref{eq:JsInnerSlope} for Sommerfeld-enhanced dark matter with $\gamma=0.7, 1.0, 1.3$.}}
    \label{fig:NFW}
\end{figure*}

\begin{figure*}[hptb]
    \centering
    \includegraphics[width=\textwidth]{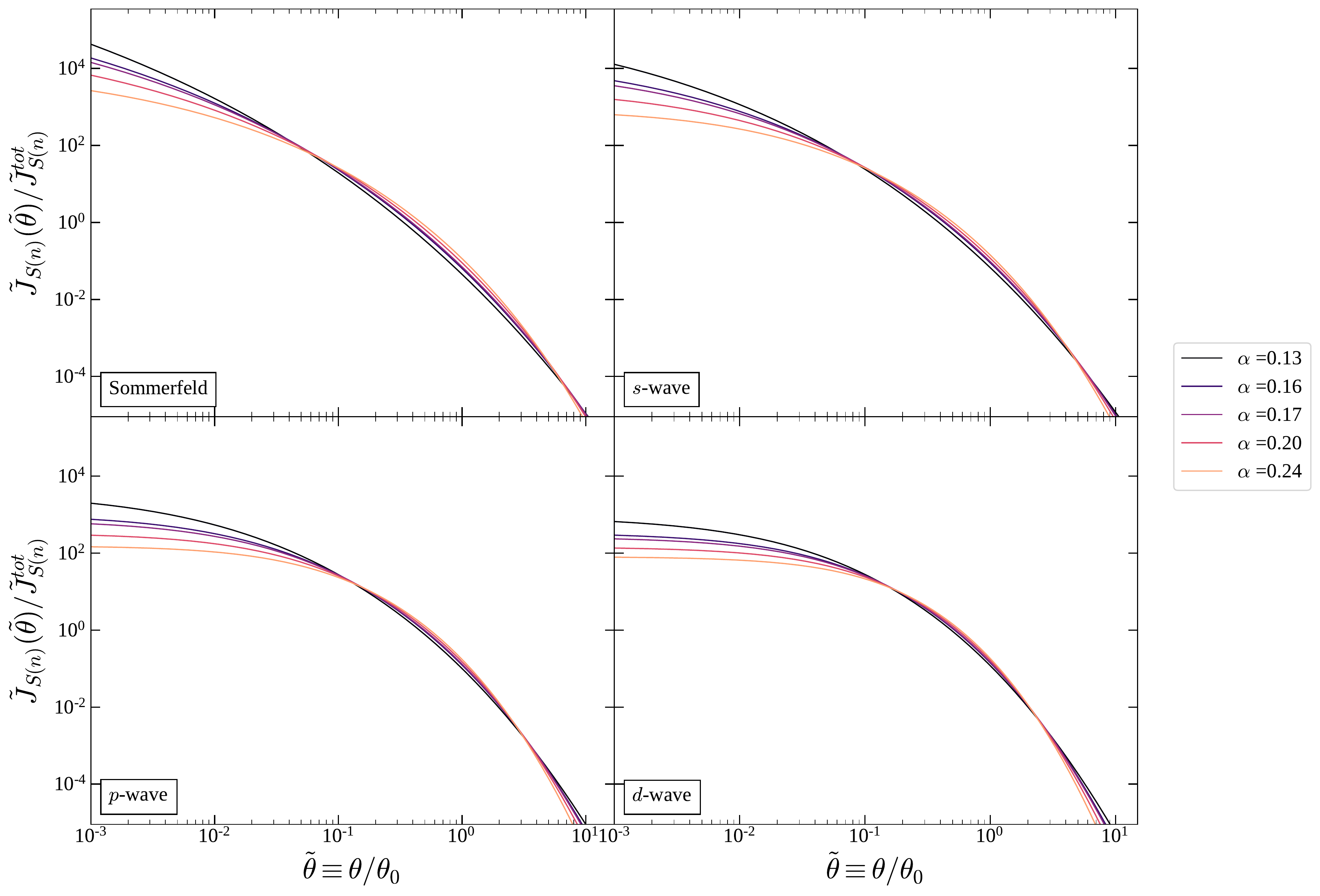}
    \caption{
    The scale-free photon angular distribution arising 
    from Sommerfeld-enhanced (upper left), $s$-wave (upper right), $p$-wave (lower left), and 
    $d$-wave (lower right) dark matter annihilation 
    in an Einasto subhalo (the profile parameter 
    $\alpha$ varies from $0.12$ to $0.25$, as labelled).}
    \label{fig:einasto}
\end{figure*}

\begin{figure}[hptb]
    \includegraphics[width=1.0\columnwidth]{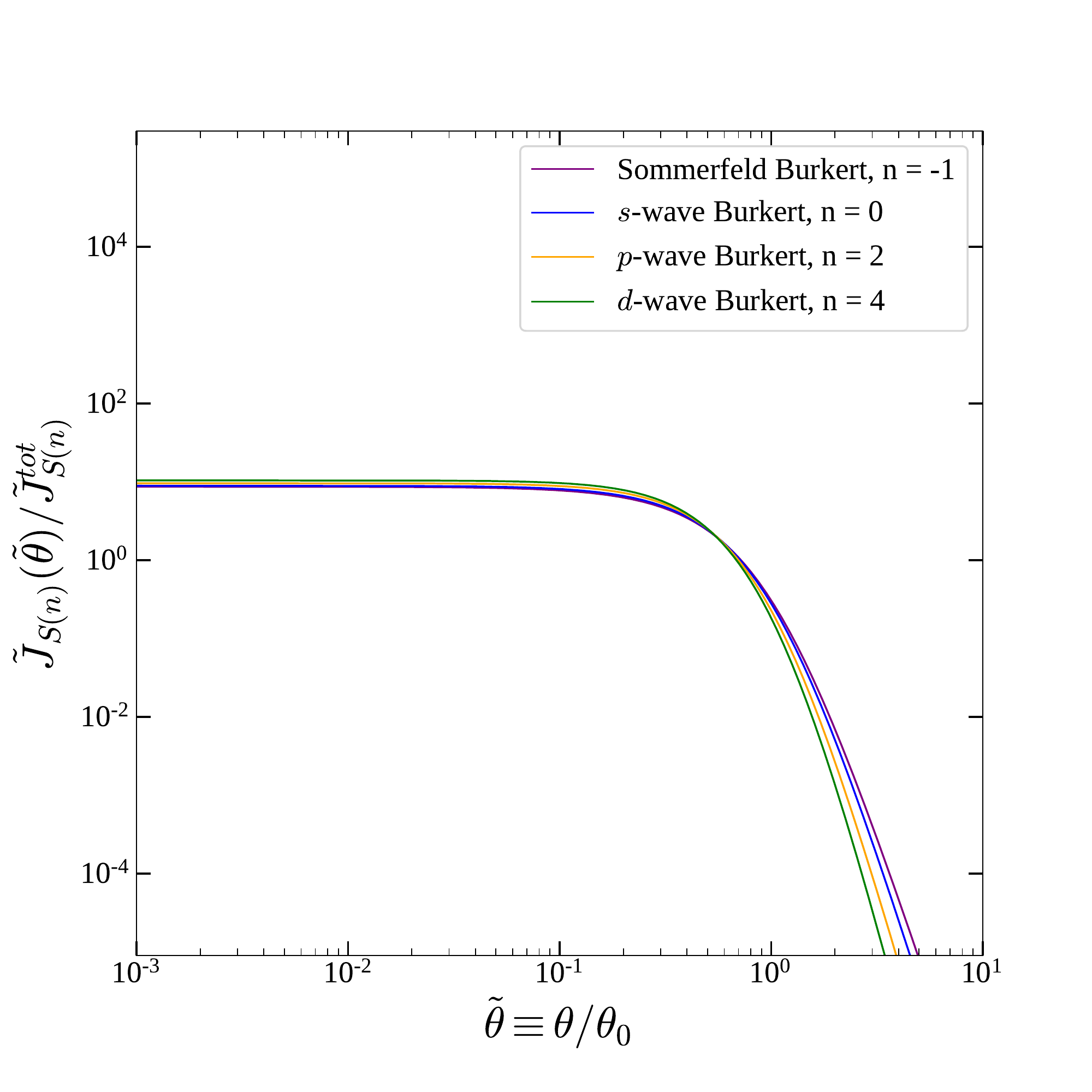}
    \caption{The scale-free photon angular distribution for the Burkert profile, with $n = -1, 0, 2, 4$, as labelled.
}
    \label{fig:burkert}
\end{figure}

\begin{figure}[hptb]
    \includegraphics[width=1.0\columnwidth]{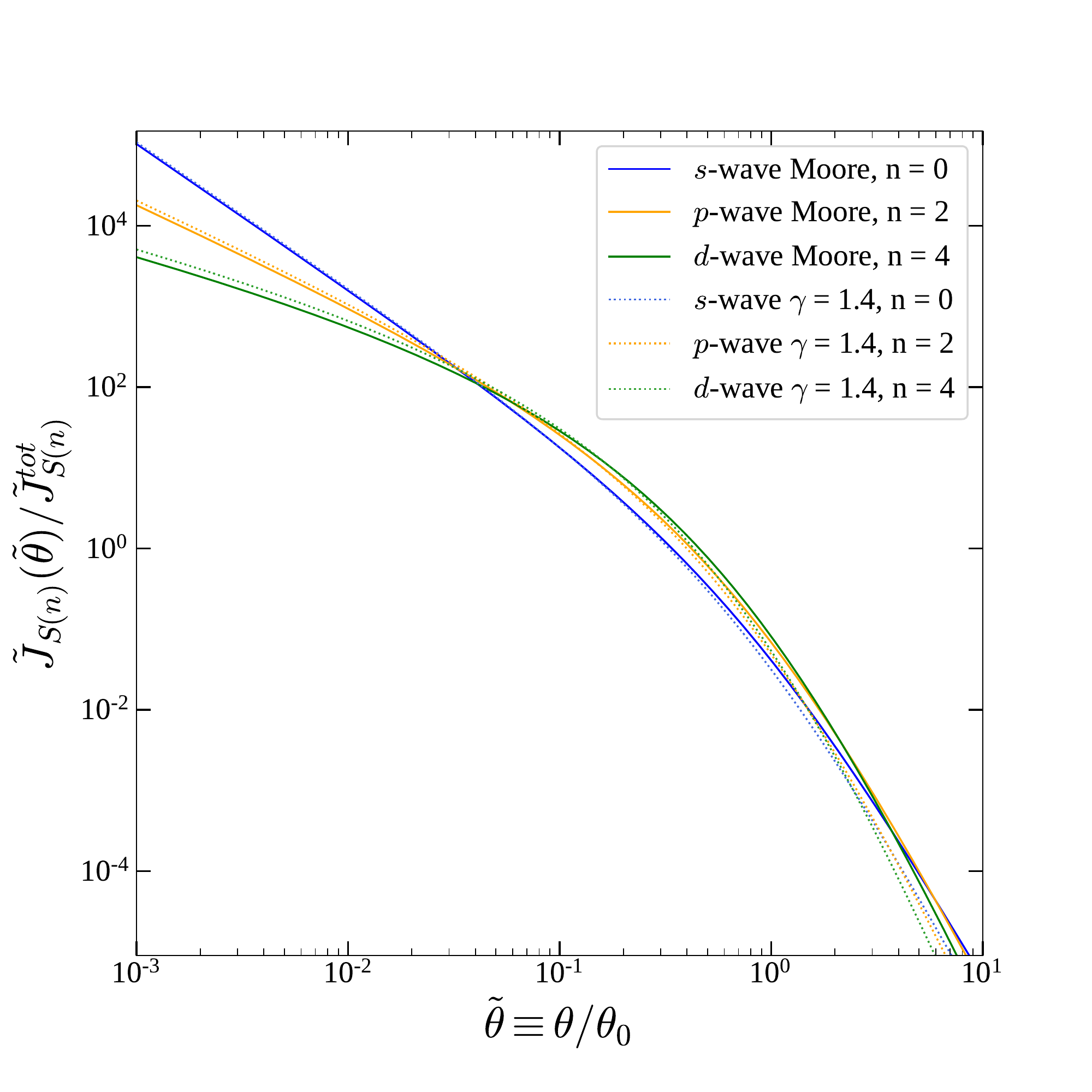}
    \caption{The scale-free photon angular distribution for the Moore profile (solid lines), with $n = 0, 2, 4$, as labelled.  For comparison, the scale-free angular distribution for generalized NFW ($\gamma = 1.4$) is also plotted (dotted lines). 
}
    \label{fig:moore}
\end{figure}

In Tables~\ref{table:Jtot} and~\ref{table:angular}, we present $\tilde J_{S(n)}^{\rm tot}$ and $\langle \theta \rangle / \theta_0$, respectively, for all of the profiles ($\tilde \rho (\tilde r)$) and choices of $n$ which we consider.
We also plot $\tilde J_{S(n)} (\tilde \theta)  / \tilde J_{S(n)}^{\rm tot}$ for all of the these profiles and choices of $n$ in Figures~\ref{fig:NFW} (generalized NFW),~\ref{fig:einasto} (Einasto),~\ref{fig:burkert} (Burkert), and ~\ref{fig:moore} (Moore).

We see that for relatively cuspy profiles, smaller values of $n$ lead to an angular distribution which is more sharply peaked at small angles.
On the other hand, we see that for a cored profile, such as Burkert, the angular distribution is largely constant at small angles, regardless of $n$.

\subsection{Inner slope limit}
To better understand the dependence of the gamma-ray angular distribution on the density profile and on the velocity-dependence of the dark matter annihilation cross section, we will consider the innermost region of the subhalo, for which $\tilde r \ll 1$.  
In this region, care must be taken during the numerical integration to achieve precise results, especially in the case of Sommerfeld-enhanced annihilation.  
\renew{The divergence near the origin requires fine-grained sampling of the integrands in order to obtain convergence of the integrals. However, the numerical accuracy of such integrals can be hard to estimate. Additionally, we cannot determine \textit{a priori} whether the integral will converge for any given model (as will be discussed for certain Sommerfeld-enhanced annihilation models later in this section.}
Fortunately, we will find that if $\tilde \rho (\tilde r)$ has power law behavior, then we can solve for $\tilde f (\tilde E)$ analytically in the inner slope region, giving us simple expressions for $P^2_n (\tilde r)$ and $\tilde J_{S(n)} (\tilde \theta)$, which can be matched to the full numerical calculation.

We may relate the density distribution to the velocity distribution using
\bea
\tilde \rho (\tilde r) &=& 4\pi \int_0^{\tilde v_{esc} (\tilde r)} d\tilde v ~
\tilde v^2 \tilde f(\tilde r, \tilde v) ,
\nonumber\\
&=& 4\sqrt{2}\pi \int_{\tilde \Phi (\tilde r)}^{\tilde \Phi (\infty)} d\tilde E ~
\tilde f (\tilde E) \sqrt{\tilde E - \tilde \Phi (\tilde r) } .
\eea

We assume that, in the inner slope region, we have $\tilde \rho (\tilde r) = \tilde \rho_0 \tilde r^{-\gamma}$, with $\gamma \new{\geq} 0$.
We then have
\bea
\tilde \Phi (\tilde r) &=& \frac{\tilde \rho_0}{(3-\gamma)(2-\gamma)} \tilde r^{2-\gamma} ,
\eea
where we adopt the convention $\tilde \Phi (0) =0$.  
Defining $x = E / \tilde \Phi (\tilde r)$, we then have
\bea
\tilde \rho_0 \tilde r^{-\gamma} &=& 
4\sqrt{2}\pi \left(\tilde \Phi (\tilde r) \right)^{3/2} 
\int_1^{\frac{\tilde \Phi (\infty)}{\tilde \Phi (\tilde r)}} dx ~
\sqrt{x-1}~ \tilde f \left(x \tilde \Phi(\tilde r)\right) .
\label{eq:rho_f_innerslope}
\nonumber\\
\eea
For $\tilde r \ll 1$ we may take $\tilde \Phi (\infty) / \tilde \Phi (\tilde r) \rightarrow \infty$, in which case the integral above depends on $\tilde r$ only through the argument of $\tilde f$.

For $\gamma > 0$, we can solve eq.~\ref{eq:rho_f_innerslope} with the ansatz $\tilde f (\tilde E) = \tilde f_0 \tilde E^\beta$, where $\beta = (\gamma - 6)/[2(2-\gamma)] < -3/2$ and 
\bea
\tilde f_0 &=& \frac{\tilde \rho_0}{4\sqrt{2} \pi} 
\left[\frac{\tilde \rho_0}{(3-\gamma)(2-\gamma)} \right]^{-(\beta + 3/2)} 
\nonumber\\
&\,& \times 
\left[\int_1^\infty dx~ x^\beta \sqrt{x-1} \right]^{-1} .
\label{eq:f0}
\eea
This matches the expression found in Ref.~\cite{Baes_2021}.
Given this expression for $\tilde f (\tilde E (\tilde r, \tilde v))$, we can perform the integral in eq.~\ref{eq:P2n}, yielding 
\bea
\tilde{P}^2_n (\tilde r \ll 1) &=& C_{\gamma, n} \tilde r^{b_n} , 
\eea
where 
$b_n ={n + \gamma (1-(6+n)/2)} $ and 
\bea
C_{\gamma, n} &=& 16\pi^2 f_0^2 \left[\frac{\tilde \rho_0}{(3-\gamma)(2-\gamma)} \right]^{2\beta + (6+n)/2}
\nonumber\\
&\,&
\times 
\int_0^\infty dy_1 \int_0^\infty dy_2~ y_1^2 y_2^2
\left[\frac{y_1^2}{2} +1 \right]^\beta  \left[\frac{y_2^2}{2} +1 \right]^\beta
\nonumber\\
&\,& \times
\left[\frac{(y_1 + y_2)^{n+2} - (|y_1 - y_2|)^{n+2}}{2(n+2)y_1 y_2} \right] .
\eea
Note, however, that this integral only converges if 
\new{
$ n < -3 - 2\beta = 2\gamma/(2-\gamma)$.}
For larger values of $n$, the dark matter annihilation rate is dominated by high velocity particles, and it is necessary to determine the velocity-distribution outside of the small $\tilde E$ regime.
But for Sommerfeld-enhanced annihilation ($n=-1$), the integral will converge for all of the cuspy slopes we consider.

Eq.~\ref{eq:Js_Jstot} then simplifies in the limit $\tilde \theta \ll 1$ to
\bea
\tilde J_{S(n)} (\tilde \theta \ll 1) &\approx& 
C_{\gamma, n} \tilde \theta^{1+b_n}  
\int_{1}^{\tilde r_0 / \tilde \theta} dx \frac{x^{b_n}}{\sqrt{1-x^{-2}}} ,
\label{eq:JsInnerSlope}
\eea
where the integral in eq.~\ref{eq:Js_Jstot} is truncated at $\tilde r_0 \leq 1$.
We assume that the power-law description of $\tilde \rho$ is accurate for $\tilde r < \tilde r_0$, and truncate the integral outside this region.
For $b_n < -1$ and $\tilde \theta \ll \tilde r_0$, the integral is insensitive to this cutoff.

For a cuspy profile, we thus have analytical expressions for the $\tilde J_{S(n)}$ at small $\tilde \theta$, and these expressions match the full expression obtained from numerical integration 
\new{(see 
Fig.~\ref{fig:NFW}, upper left panel)}.
It is interesting to note that the exponent $b_n$ exhibits a degeneracy between $\gamma$ and $n$.
Thus, for example, the power law behavior of $\tilde J_{S(n)}$ for the case of Sommerfeld-enhanced annihilation ($n=-1$) and a pure NFW profile ($\gamma = 1$) is identical to that of $s$-wave annihilation ($n=0$) for a generalized NFW profile with $\gamma = 1.25$.
However, the normalization coefficients $C_{\gamma, n}$ are different.
This implies that, for a cuspy profile, a detailed analysis of the angular distribution at both small angles and intermediate angles is in principle sufficient to resolve the velocity-dependence of dark matter annihilation.

\begin{figure}[hptb]
    \includegraphics[width=0.95\columnwidth]{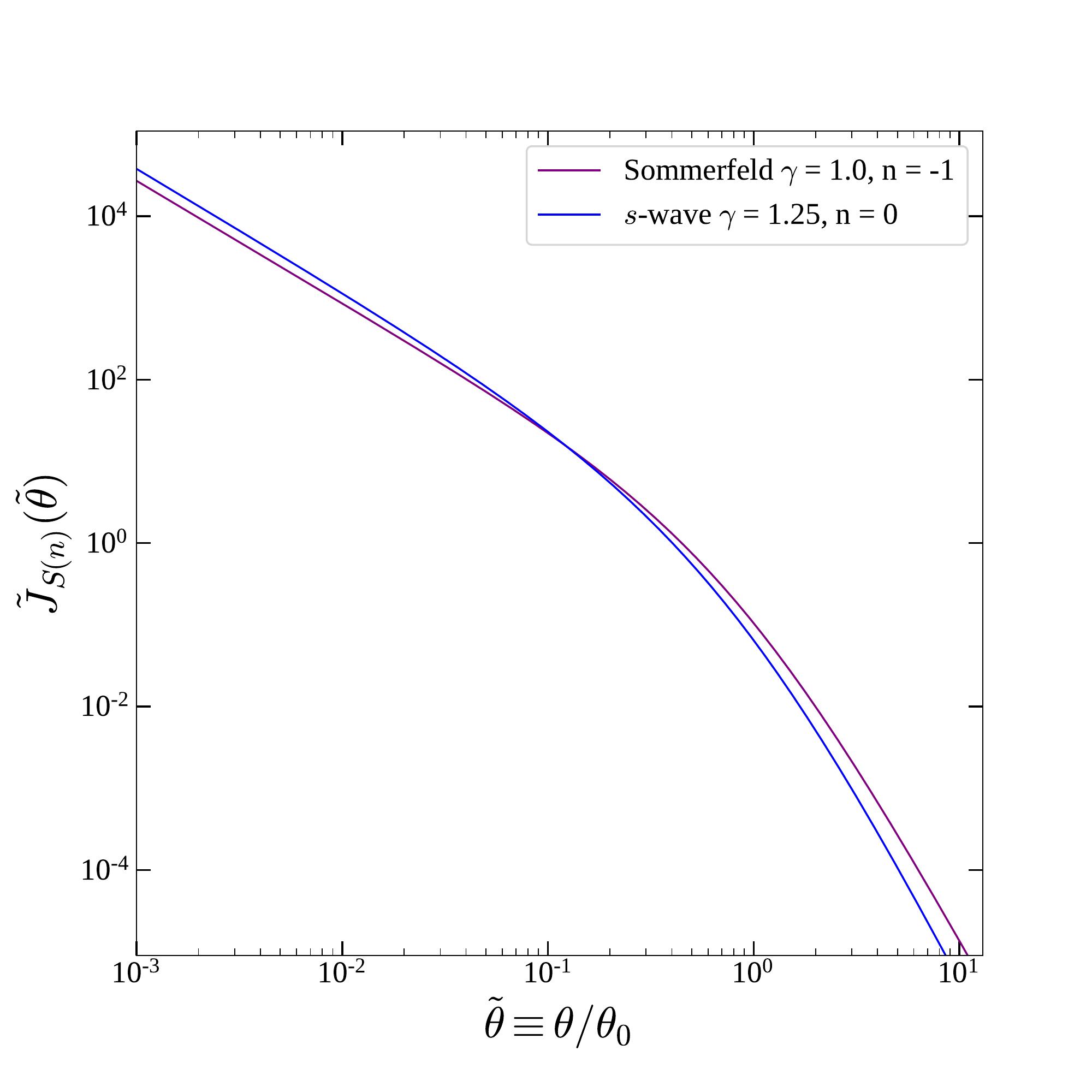}
    \caption{The scale-free photon angular distribution for a generalized NFW profile, with either $\gamma = 1.25$, $n=0$ (blue) or $\gamma = 1.0$, 
    $n=-1$ (purple).}
    \label{fig:125}
\end{figure}

To illustrate this point, in Fig.~\ref{fig:125} we plot $\tilde J_{S(n)}(\tilde \theta) / \tilde J_{S(n)}^{tot}$ for two generalized NFW profiles, $\gamma = 1$ ($n=-1$) and $\gamma = 1.25$ ($n=0$).  
This figure confirms our analytical result; both of these models yield angular distributions which exhibit the same behavior at small angles.
But they differ at larger angles, implying that with sufficient data and angular resolution, it is in principle possible to determine the velocity-dependence of the annihilation cross section.  Indeed, for 
$\gamma = 1.25$, $n=0$, we find $\langle \theta \rangle / \theta_0 = 0.21$, which is significantly smaller than 
the value found for $\gamma = 1.0$, $n=-1$ 
($\langle \theta \rangle / \theta_0 = 0.32$).  
\new{This result is to be expected, since the $\gamma = 1.25$, $n=0$ model 
has a much more cuspy profile than the $\gamma = 1.0$, $n=-1$ model.}
\renew{Moreover, both profiles illustrated in Fig.~\ref{fig:125} have a density which falls of as $r^{-3}$ at 
large distance.  If the profile were made less steep at large 
distances (in order for the angular distribution to fall off less 
rapidly), the mass of the halo would grow as a power law with distance.}
Thus, 
if the slope of angular dependence in the innermost 
region can be determined, then the scale at which that 
power law behavior cuts off is sufficient to distinguish $s$-wave annihilation from Sommerfeld-enhanced  
annihilation, with Sommerfeld-enhanced annihilation 
producing a more extended angular distribution.
\renew{Although we have plotted the angular distributions  
in terms of $\tilde \theta = \theta / \theta_0$, this result 
does not depend on one's ability to determine $r_S$ experimentally.  A rescaling of $r_s$ (or, equivalently, $\theta_0$) would amount 
to a shift of one of the curves plotted in Fig.~\ref{fig:125}, but not a change in its shape.}

In a similar vein, we have compared the 
angular distribution 
for the Moore profile and generalized NFW profile 
($\gamma = 1.4$) in Figure~\ref{fig:moore}.  
Both profiles have the same inner slope, but the
Moore profile yields more extended emission.  This 
result is echoed in Table~\ref{table:angular}, where 
we see that $\langle \theta \rangle / \theta_0$ is 
about $\sim 20\%$ larger for a Moore profile than 
for generalized NFW with $\gamma = 1.4$, for 
$n = 0, 2, 4$.

\begin{figure*}[hptb]
    \includegraphics[width=0.95\textwidth]{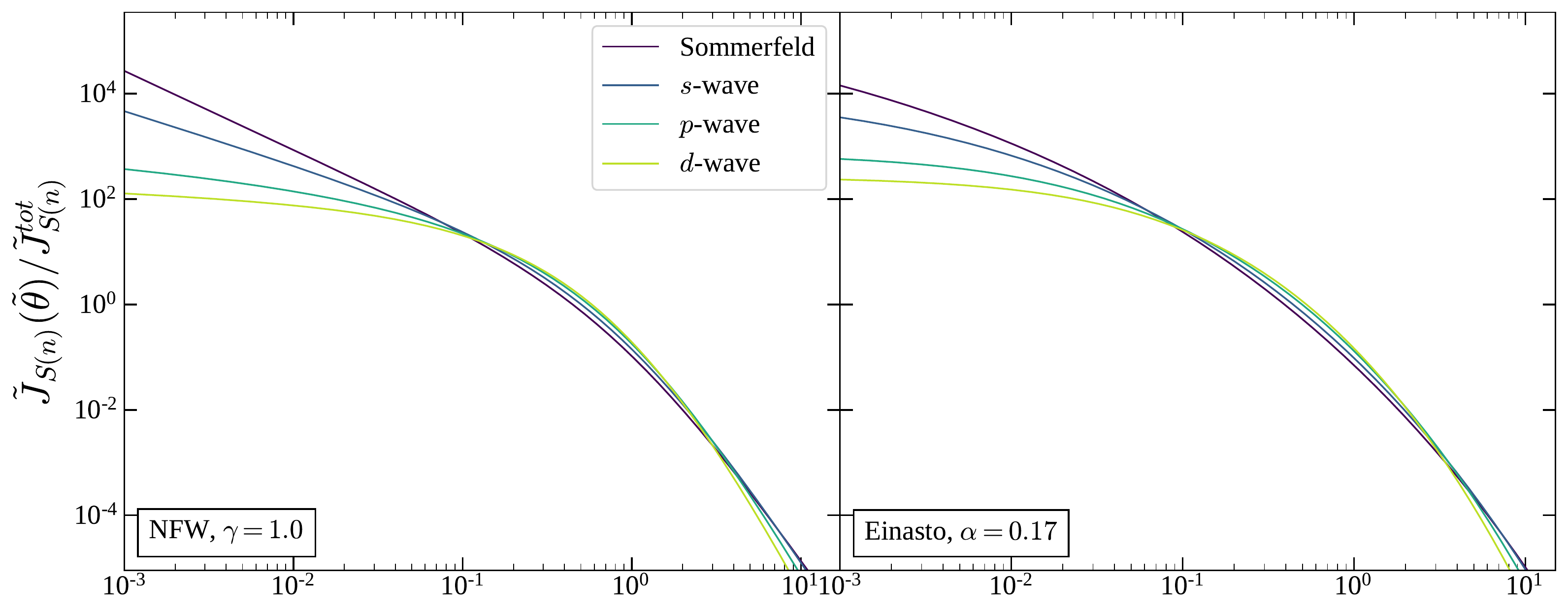}
    \caption{Comparisons of different velocity-dependent models with the same DM profile. }
    \label{fig:n_compare}
\end{figure*}

\new{In Fig.~\ref{fig:n_compare}, we supplement the values of $\langle\theta\rangle/\theta_0$ by illustrating the differences in the angular spread of the annihilation of a given DM profile for different velocity-dependent models.
For the cuspy profiles, Sommerfeld emission dominates near the center and at larger angles but is the smallest in between.
On the other hand, $d$-wave emission is smallest near the center and at larger angles but dominates in between.
Quantitatively, we can see from Table~\ref{table:angular} that for the cuspy profiles $\langle\theta\rangle/\theta_0$ increases with increasing $n$.
}

Interestingly, we find that, for Sommerfeld-enhanced annihilation ($n=-1$), we have $b_{n=-1} < -3$  for $\gamma > 4/3$.
For $b_n < -3$, the integral for $\tilde J_{S(n)}^{tot}$ diverges at small $\tilde \theta$.
This implies that for a profile, such as Moore, with $\gamma > 4/3$,  our treatment of Sommerfeld-enhanced annihilation has been inconsistent.
In particular, we have implicitly assumed that dark matter annihilation does not deplete the dark matter density significantly, which may not be the case.
Moreover, the $1/v$ Sommerfeld-enhancement of the annihilation cross section is cut off at a velocity-scale which depends on the mediator 
mass~\cite{ArkaniHamed:2008qn}, and we have assumed that this cutoff is at a velocity small enough to be irrelevant.

It is also interesting to note that, for cuspy 
profiles
$\tilde J_{S(n=2)}^{tot}$ tends to be  
significantly smaller than $\tilde J_{S(n=0)}^{tot}$, 
while $\tilde J_{S(n=4)}^{tot}$ is only a slightly 
smaller than 
$\tilde J_{S(n=2)}^{tot}$.  This may seem 
counter-intuitive, since the integrals which determine 
$P_n^2$ have integrands which scale as powers of 
$\tilde v^n$.  But as we have seen, for larger 
$n$, $P_n^2$ becomes more sensitive to the high-velocity tail of particles which are not 
confined to core.  As a result, we find 
$\langle \tilde v^4 \rangle \gg 
(\langle \tilde v^2 \rangle )^2$.

\subsection{Cored profile}

The situation is somewhat different for a cored profile.
For the Burkert profile, which exhibits a core, the differences in the angular distribution arising from $n = -1, 0, 2$ or $4$ are much smaller.  In particular, 
the angular distribution is flat at small angles, 
regardless of $n$.  
This implies that morphology of the photon signal carries less information regarding the velocity-dependence of dark matter annihilation. 

We can again understand this behavior by considering an analytic approximation.
Let us approximate the cored profile with $\tilde \rho (\tilde r) = \tilde \rho_0$ for $\tilde r < 1$, and assume the density vanishes rapidly for $\tilde r > 1$.
For $\tilde r < 1$ we then have $\tilde \Phi (r) = (\tilde \rho_0 /6) \tilde r^2$, and Eq.~\ref{eq:rho_f_innerslope} can be rewritten as 
\bea
\tilde \rho_0 &=& 
4\sqrt{2} \pi \left[\frac{\tilde \rho_0 \tilde r^{2}}{6} \right]^{3/2}
\int_1^{\tilde r^{-2}} dx~ \sqrt{x-1} \times
\tilde f\left(x  \frac{\tilde \rho_0 \tilde r^{2}}{6} \right),
\nonumber\\
\eea
for small $\tilde r$,
where we have made the approximation that particles do not explore the region outside the core.
In this case, one cannot find a power-law solution for $\tilde f$ while taking the upper limit of integration to infinity, as the integral would not converge.
Instead, this equation can be solved 
\new{for $\tilde r \ll 1$} 
by taking $\tilde f = (9\sqrt{3} / 4\pi) \tilde \rho_0^{-1/2}$.

We thus see that, for a cored profile, the velocity distribution is independent of $\tilde E$ for paths confined to the innermost region.  
This implies that, for $\tilde r \ll 1$, $\tilde f$,  and thus $P^2_n$, are independent of $\tilde r$.
If the velocity distribution is independent of $\tilde r$, the angular distribution of the gamma-ray signal cannot depend on $n$, since the effects of velocity-suppression do not depend on the distance from the center of the subhalo.
Indeed, we can confirm this result by noting that, for a cored profile, since $P^2_n$ is independent of $\tilde r$ at small $\tilde r$ for all $n$, we can rewrite Eq.~\ref{eq:JsInnerSlope} as 
\bea
\tilde J_{S(n)}^{cored} (\tilde \theta) &\propto& 
\tilde \theta \int_1^{\tilde r_0 / \tilde \theta} dx~[1-x^{-2}]^{-1/2} .
\eea
But in this case, we cannot ignore the upper limit of integration, and we find that $\tilde J_{S(n)}^{cored} (\tilde \theta)$ becomes independent of $\tilde \theta$ at small angle.

This result 
matches what is found from a complete numerical 
calculation for the Burkert profile.  More generally, 
we see from Table~\ref{table:angular} that, 
as profiles become more cored, the difference in $\langle \theta \rangle / 
\theta_0$ between the $n=-1, 0, 2$ and $4$ become 
smaller.  
The above argument suggests 
that the degeneracy of all four cases is only broken 
by the behavior of the profile at larger $\tilde r$, 
as one leaves the core.

For a Burkert profile, $\langle \theta \rangle / 
\theta_0$ tends to decrease as $n$ increases.  This 
behavior can be readily understood, because 
annihilation at large angles is dominated by particles 
which are far from the core.  As particles get 
farther from the core, the escape velocity (which 
is the largest allowed velocity for a bound particle) 
decreases, suppressing annihilation for larger 
$n$.  But interestingly, $\langle \theta \rangle / 
\theta_0$ tends to increase with $n$ for the case 
of generalized NFW.  The suppression of annihilation 
far from the core with larger $n$ still occurs in 
this case.  But there is an additional effect; 
$P_n^2 (r)$ has a less steep slope in the inner 
region for large $n$.  Thus, for cuspy profiles, 
as $n$ increases, the angular distribution is 
suppressed both at large and very small angles, with 
the overall effect being to increase the average 
angular size of emission.  For a cored profile like 
Burkert, on the other hand, the second effect is 
not present, as the angular distribution 
in the inner slope region is flat for any $n$.

\section{Conclusion}
\label{sec:conclusion}

We have determined the effective $J$-factor for the cases of $s$-wave, $p$-wave, $d$-wave and Sommerfeld-enhanced (in the Coulomb limit) dark matter annihilation for a variety of dark matter profiles, including generalized NFW, Einasto, Burkert, and Moore.
We have assumed that the dark matter velocity distribution is spherically-symmetric and isotropic, and have recovered the velocity distribution from the density distribution by numerically solving the Eddington inversion equation.
If the density-profile is power-law in the inner slope region, then the velocity-distribution in the inner slope region can be determined analytically, yielding results which match the full numerical calculation.

\new{We have found that, for a large class of profiles,  
the angular dependence of the photon flux at small angles is completely determined by the steepness of the cusp and the power-law velocity dependence.
Although there is a degeneracy between these two quantities in the angular distribution at small angles, this degeneracy is broken at larger angles.
}

\new{
For a cored profile, on the other hand, 
the velocity distribution is largely independent of position.
Thus, although the velocity-dependence of the annihilation cross section will affect the overall rate of dark matter annihilation, it will not affect the distribution within the core.  Instead, the effect of the velocity-dependence on the photon angular distribution is largely determined by what happens at the edge of the core.  
}

Our analysis has focused on the magnitude and angular distribution of the dark matter signal.
We have not considered astrophysical backgrounds, or the angular resolution of a realistic detector.
It would be interesting to apply these results to a particular instrument in development, to determine the specifications needed to distinguish the velocity-dependence of a potential signal in practice. 
For a cuspy profile, it is apparent from Figure~\ref{fig:NFW} that, to resolve the power-law angular slope dependence
of the inner slope region, one would need an angular 
resolution of better than $1/10$ of the angle subtended 
by the scale radius.

Interestingly, we have found that if the dark matter density profile has a power-law steeper than $\gamma = 4/3$ (an example is the Moore profile), then the rate of Sommerfeld-enhanced annihilation in the Coulomb limit diverges at the core.
In a specific particle physics model, one expects that the $1/v$ Sommerfeld-enhancement in the Coulomb limit will not be valid at arbitrarily small velocities, unless the particle mediating dark matter self-interactions is truly massless.
It is often assumed that this cut off occurs at velocities which are negligible, but if the profile is steep enough, then this effect cannot be ignored.  
Moreover, if the dark matter annihilation rate at the core is sufficiently large, then the effect of annihilation on the dark matter distribution also cannot be ignored.
It would be interesting to consider Sommerfeld-enhanced annihilation in the very cuspy limit in more detail.

As we have seen, one would need excellent angular 
resolution to robustly distinguish the dark matter 
velocity-dependence of a single dark matter subhalo
(for recent work on determining the velocity-dependence using an ensemble of subhalos, see, for example,~\cite{Baxter:2021pzi,Runburg:2021pwh}).
The Galactic Center is a larger target, and it would 
be interesting to perform a similar analysis for that 
case.  One important difference, in that case, is 
that there is a large baryonic contribution to the 
gravitational potential, which would affect the 
dark matter velocity distribution.

{\bf Acknowledgements}

We are grateful to Andrew B.~Pace and Louis E.~Strigari for useful discussions.
The work of BB and VL is supported in part by the Undergraduate Research 
Opportunities Program, Office of the Vice Provost for Research and Scholarship 
(OVPRS) at the University of Hawai`i at Mānoa.
The work of JK is supported in part by DOE grant DE-SC0010504.
The work of JR is supported by NSF grant AST-1934744.

\bibliography{dSph_Jfactors}

\end{document}